\documentclass{book}
\usepackage{maxent,xspace}

\def\bm#1{\mbox{\boldmath $#1$}}

\def\bb{{\bm b}}

\def\rb{{\bm r}}

\def\ub{{\bm u}}

\def\xb{{\bm x}}
\def\yb{{\bm y}}
\def\zb{{\bm z}}

\def\Ab{{\bm A}}

\def\Hb{{\bm H}}
\def\Ib{{\bm I}}

\def\Pb{{\bm P}}

\def\Rb{{\bm R}}

\def\thetab{\bm{\theta}}

\def\lambdab{\bm{\lambda}}
\def\phib{\bm{\phi}}

\def\wh#1{\widehat{#1}}
\def\ra{\rightarrow}

\def\lra{\longrightarrow}

\def\Lra{\Longrightarrow}

\def\d#1{\,\mbox{d}#1}

\def\disp#1{{\displaystyle #1}}

\def\intd{\int\kern-.8em\int}
\def\intt{\int\kern-.8em\int\kern-.8em\int}
\def\intg{\int\kern-1.1em\int}
\def\expf#1{\exp\left[ {#1} \right]}

\def\argmins#1#2{\mbox{arg}\min_{#1}\left\{{#2}\right\}}
\def\argmaxs#1#2{\mbox{arg}\max_{#1}\left\{{#2}\right\}}

\def\esp#1{\mbox{E}\left\{ #1 \right\}}
\def\espx#1#2{\mbox{E}_{#1}\left\{ #2 \right\}}

\def\uncatcodespecials{\def\do##1{\catcode`##1=12 }\dospecials}

\newcount\lineno
\def\setupverbatim{\tt \lineno=0
 \obeylines \uncatcodespecials \obeyspaces
 \everypar{\advance\lineno by1 \llap{\sevenrm\the\lineno\ \ }}}
{\obeyspaces\global\let =\ }

\def\ER{\mbox{I\kern-.25em R}}
\def\EC{\mbox{C\kern-.8em C}}
\def\EZ{\mbox{Z\kern-.55em Z}}
\def\EN{\mbox{N\kern-.8em N}}

\def\beqnarr#1&#2&#3\\#4&#5&#6\eeqnarr{
    \left\{
           \begin{array}{lcl}
            {\displaystyle #1} & #2 & {\displaystyle #3} \\ 
            {\displaystyle #4} & #5 & {\displaystyle #6} 
           \end{array}
    \right. }

\def\pyx{p(\yb|\xb)}
\def\pxy{p(\xb|\yb)}

\def\ie{{\em i.e.}}

\title{A SCALE INVARIANT BAYESIAN METHOD TO SOLVE LINEAR INVERSE PROBLEMS} 

\author{Ali Mohammad-Djafari and J\'er\^ome Idier \\
Laboratoire des Signaux et Syst\`emes (CNRS-ESE-UPS) \\ 
\'Ecole Sup\'erieure d'\'Electricit\'e, \\ 
Plateau de Moulon, 91192 Gif-sur-Yvette C\'edex, France}

\begin{document}
\maketitle
\thispagestyle{empty}


\begin{abstract}
In this paper we propose a new Bayesian estimation method to solve linear
inverse problems in signal and image restoration and reconstruction problems 
which has the property to be scale invariant. 
In general, Bayesian estimators are {\em nonlinear} functions of the observed
data. The only exception is the Gaussian case. When dealing with linear 
inverse problems the linearity is sometimes a too strong property, while 
{\em scale invariance} often remains a desirable property.  
As everybody knows one of the main difficulties with using the Bayesian 
approach in real applications is the assignment of the direct (prior)
probability laws before applying the Bayes' rule.  We discuss here how to
choose prior laws to obtain scale invariant Bayesian estimators. 
In this paper we discuss and propose a familly of generalized exponential
probability distributions functions for the direct probabilities 
(the prior $p(\xb)$ and the likelihood $p(\yb|\xb)$), for which the posterior
$p(\xb|\yb)$, and, consequently, the main posterior estimators are scale
invariant.  Among many properties, generalized exponential can be considered
as the maximum entropy probability distributions subject to the knowledge of a
finite set of expectation values of some knwon functions. 
\end{abstract}

\section{Introduction}
We address a class of linear inverse problems arising in signal and image
reconstruction and restoration problems which is to solve integral equations 
of the form: 
\begin{equation} \label{eq:1}
g_{ij}=\intg_D f(\rb') \, h_{ij}(\rb') \d{\rb'} + b_{ij},\quad i,j=1,\cdots,M,
\end{equation}
where $\rb'\in \ER^2$, $f(\rb')$ is the object (image reconstruction problems) 
or the original image (image restoration problems), $g_{ij}$  are the measured 
data (the projections in image reconstruction or the degraded image in image
restoration problems),  $b_{ij}$ are the measurement noise samples and
$h_{ij}(\rb')$ are known functions which depend only on the measurement
system. To show the generality of this relation, we give in the following 
some applications we are interested in:  
\begin{itemize}
\item  Image restoration:
\[
g(x_i,y_j)=\intd_D f(x',y') h(x_i-x',y_j-y')\d{x'}\d{y'}+b(x_i,y_j)\quad, 
\begin{array}{l} i=1,\cdots,N \\ j=1,\cdots,M \end{array}, 
\]
where $g(x_i,y_j)$ are the observed degraded image pixels and $h(x,y)$ is 
the point spread function (PSF) of the measurement system.

\item X-ray computed tomography (CT):
\[
g(r_i,\phi_j)=\intd_D f(x,y)
\delta(r_i-x\cos\phi_i-y\sin\phi_i)\d{x}\d{y}+b(r_i,\phi_j)
\quad, \begin{array}{l} i=1,\cdots,N \\ j=1,\cdots,M \end{array},
\]
where $g(r_i,\phi_j)$ are the projections along the axis 
$r_i=x \cos\phi_i - y \sin\phi_i$,  having the angle $\phi_j$, and which can
be considered as the samples of the Radon transform (RT) of the  object
function $f(x,y)$. 

\item Fourier Synthesis in radio astronomy, in SAR imaging and in
diffracted wave tomographic imaging systems: 
\[
g(u_j,v_j)=\intd_D f(x,y) \, \expf{-j(u_j x+v_j y)}\d{x}\d{y}+b(u_j,v_j),
\quad j=1,\cdots,M,
\]
where $\ub_j=(u_j,v_j)$ is a radial direction and $g(u_j,v_j)$ are the 
samples of the complex valued visibility function of the sky in radio
astronomy or the Fourier transform of the  measured signal in SAR imaging. 
\end{itemize}
Other examples can be found in \cite{AMD87a,AMD88a,AMD90b,AMD93a,AMD93b}. 

In all these applications we have to solve the following ill-posed problem: 
how to estimate the function $f(x,y)$ from some finite set of measured data 
which may also be noisy, because there is no experimental measurement device,
even the most elaborate, which could be entirely free from uncertainty, the
simplest example being the finite precision of the measurements. 

The numerical solution of these equations needs a discretization procedure 
which can be done by a quadrature method. The linear system of equations
resulting from the discretization of an ill-posed problem is, in general, very
ill-conditioned if not singular. So the problem is to find a unique and stable
solution for this linear system. The general methods which permit us to find a
unique and stable solution to an ill-posed problem by introducing an {\em a priori}
information on the solution are called regularization . The {\em a priori}
information can be either in a deterministic form (positivity) or in a
stochastic form (some constraints on the probability density functions). 

When discretized, these problems can be described by the following: 
\begin{quote}
``Estimate a vector of the parameters $\xb\in \ER^n$  (pixel intensities in
an image for example) given a vector of measurements $\yb\in \ER^m$ 
(representing, for example, either a degraded image pixel values in
restoration problems or the projections values in reconstruction problems) 
and a linear transformation $\Ab$ relating them by: 
\begin{equation} \label{eqdiscret}
\yb=\Ab\xb + \bb,
\end{equation}
where $\bb$ represents the discretization errors and the measurement noise 
which is supposed to be zero-mean and additive.'' 
\end{quote}

In this paper we propose to use the Bayesian approach to find a regularized 
solution to this problem. Noting that the Bayesian theory only gives us a
framework for the formulation of the inverse problem, not a solution of it. 
The main difficulty is, in general, before the application of the Bayes'
formula, \ie; how to formulate appropriately the problem and how to assign the
direct probabilities. Keeping this fact in mind, we propose the following
organization to this paper:  In section 2. we give a brief description of the
Bayesian approach with detail calculations of the solution in the special case
of Gaussian laws. In section 3. we discuss about the {\em scale invariance}
property and propose a familly of prior probability density functions ($pdf$)
which insure this property for the solution.  Finally, in section 4., we present
some special cases and give detailed calculations for the solution.

\section{General Bayesian approach}
A general Bayesian approach involves the following steps:
\begin{itemize}
\item Assign a prior probability law $p(\xb)$ to the unknown parameter
      to translate our incomplete {\em a priori} information (prior beliefs) about 
      these parameters;  

\item Assign a direct probability law to the measured data $\pyx$ to 
      translate the lack of total precision and the inevitable existence 
      of the measurement noise; 

\item Use the Bayes' rule to calculate the posterior law $\pxy$ of
       the unknown parameters; 

\item Define a decision rule to give values $\wh{\xb}$ to these parameters. 
\end{itemize}
To illustrate the whole procedure, let us to consider an example; the 
Gaussian case. If we suppose that what we know about the unknown input $\xb$
is its mean $\esp{\xb}=\xb_0$ and its covariance matrix
$\esp{(\xb-\xb_0)(\xb-\xb_0)^t}=\Rb_x=\sigma_x^2 \Pb$,  and what we
know about the measurement noise $\bb$ is also its covariance matrix 
$\esp{\bb\bb^t}=\Rb_b=\sigma_b^2\Ib$,  then we can use the maximum entropy
principle to assign: 
\begin{equation} \label{px} 
p(\xb)\propto \expf{-\frac{1}{2} (\xb-\xb_0)^t{\Rb_x}^{-1}(\xb-\xb_0)},  
\end{equation}
and
\begin{equation} \label{pyx}
\pyx\propto \expf{-\frac{1}{2} (\yb-\Ab\xb)^t {\Rb_b}^{-1} (\yb-\Ab\xb)}.
\end{equation}
Now we can use the Bayes' rule to find:
\begin{equation} \label{Bayes}
\pxy  \propto \pyx \, p(\xb),
\end{equation}
and use, for example, the maximum a posteriori (MAP) estimation rule to give 
a solution to the problem, \ie; 
\begin{equation}
\wh{\xb} =\argmaxs{\xb}{\pxy},
\end{equation}
Other estimators are possible. In fact, all we want to know is resumed in the 
posterior law. In general, one can construct a bayesian estimator by defining
a cost (or utility) function $C(\wh{\xb},\xb)$ and by minimizing its mean value
\[
\wh{\xb}=\argmins{\zb}{\espx{X|Y}{C(\zb,\xb)}} 
     =\argmins{\zb}{\intg C(\zb,\xb)p(\xb|\yb)\d{\xb}}.  
\]
The two classical estimators:
\begin{itemize}
\item Posterior mean (PM): \quad  
$\disp{\wh{\xb}=\espx{X|Y}{\xb}=\intg \xb \, p(\xb|\yb) \d{\xb}},$ 

is obtained when defining $C(\wh{\xb},\xb)=(\wh{\xb}-\xb)^t(\wh{\xb}-\xb)$, and  
 
\item Maximum {\em a posteriori} (MAP): \quad 
$\disp{\wh{\xb}=\argmaxs{\xb}{p(\xb|\yb)}},$ 

is obtained when defining $C(\wh{\xb},\xb)=1-\delta(\wh{\xb}-\xb)$. 
\end{itemize}

Now, let us go a little further inside the calculations. 
Replacing (\ref{px}), and (\ref{pyx}) in (\ref{Bayes}),  we calculate the
posterior law:  \[
\pxy \propto \expf{-\frac{1}{2\sigma_b^2} J(\xb)},
\hbox{~~with~}
J(\xb)=(\yb-\Ab\xb)^t(\yb-\Ab\xb)
+\lambda(\xb-\xb_0)^t{\Pb}^{-1}(\yb-\xb_0),
\]
where $\lambda={\sigma_b^2}/{\sigma_x^2}$. 
The posterior is then also a Gaussian. 
We can know use any decision rule to obtain a
solution. For example the maximum a posteriori (MAP) solution is obtained by:
\begin{equation} \label{eq:xbh} 
\wh{\xb} =\argmaxs{\xb}{\pxy}=\argmins{\xb}{J(\xb)}.
\end{equation}
Note that in this special Gaussian case both estimators, \ie;  the posterior
mean (PM)  and the MAP estimators are the same:
\begin{equation}
\wh{\xb} =\espx{X|Y}{\xb}=\argmaxs{\xb}{\pxy}
\end{equation}
and the minimization of the criterion $J(\xb)$  which can also be written in
the form:  
\begin{equation}
J(\xb)=||\yb-\Ab \xb||^2 + \lambda ||\xb-\xb_0||_{\Pb}^2
\end{equation}
can be considered as a regularization procedure to the inverse problem 
(\ref{eqdiscret}). Indeed, the Bayesian approach will give us here a new
interpretation of the regularization parameter in terms of the signal to noise
ratio, \ie; $\lambda= \sigma_b^2 /  \sigma_x^2$.

$J(\xb)$ is a quadratic function of $\xb$. The solution $\wh{\xb}$ is then a
linear function of the data $\yb$. This is due to the fact that the problem 
is linear and all the probability laws are Gaussian.  
Excepted this case, in general, the Bayesian estimators are not linear
functions of the observations $\yb$. 
However, we may not need that the solution be a linear function of the
data $\yb$, but  the {\em scale invariance} is the minimum property which 
is often needed.  

\section{Scale invariant Bayesian estimators}
What we are proposing in this paper is to study in what conditions we can
obtain estimators who are scale invariant. Note that {\em linearity} is the
combination of  
\begin{eqnarray*}
\hbox{\em additivity:}
&&\left\{\begin{array}{l}\yb_1 \mapsto \wh{\xb}_1,\\  \yb_2 \mapsto \wh{\xb}_2\end{array}\right. 
\Lra   
\yb_1 + \yb_2 \mapsto \wh{\xb}_1 + \wh{\xb}_2, 
\\ 
\hbox{and}\hspace{2in} && ~
\\ 
\hbox{\em scale invariance:}
&& \yb_1 \mapsto \wh{\xb}_1
\Lra
\forall k>0, \,  k \yb_1 \mapsto k \wh{\xb}_1. 
\end{eqnarray*}
In a linear inverse problem what is often necessary is that the solution be
scale invariant. As we have seen in the last section when all the probability
laws are Gaussian then the Bayesian estimators are linear functions of the
data, so that the methods based on this assumption have not to take care
about the scale of the measured data.  The Gaussian assumption is very
restrictive. On the other hand, more general priors yield  the Bayesian
estimators which are nonlinear functions of data, so the result of the
inversion method depend on the absolute values of the measured data.  
In other words, two users of the method using two different scale factors 
would not get the same results, even rescaled:

\medskip
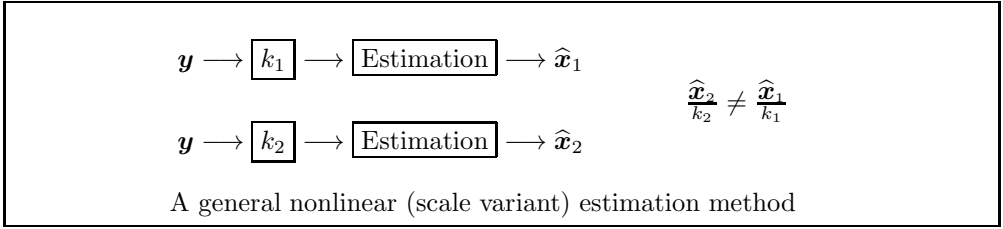
\fbox{\hbox to 13cm{\vbox{
\[
\begin{array}{ll}
\yb\lra\fbox{$k_1$}\lra\fbox{Estimation}\lra\wh{\xb}_1 \\ 
    &\hspace{1cm} \frac{\wh{\xb}_2}{k_2} \not= \frac{\wh{\xb}_1}{k_1}  \\ 
\yb\lra\fbox{$k_2$}\lra\fbox{Estimation}\lra\wh{\xb}_2
\end{array}
\]
\centerline{A general nonlinear (scale variant) estimation method}
}}}

\medskip\noindent  
What we want to specify in this paper is a family of probability laws for
which these estimators are scale invariant. So the user of the inversion
method can process the data without worrying about rescaling them to an
arbitrary level and two users of the method at two different scales will
obtain the proportional results:

\medskip
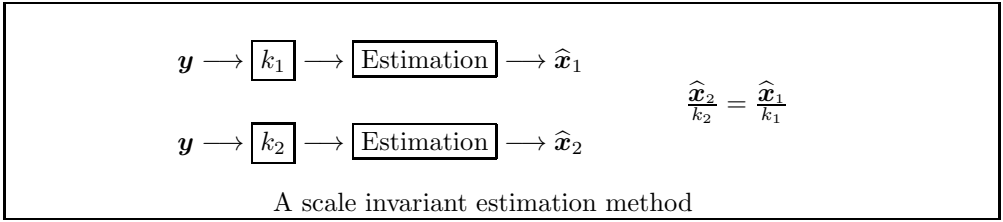
\fbox{\hbox to 13cm{\vbox{
\[
\begin{array}{ll}
\yb\lra\fbox{$k_1$}\lra\fbox{Estimation}\lra\wh{\xb}_1 \\ 
 &\hspace{1cm} \frac{\wh{\xb}_2}{k_2} = \frac{\wh{\xb}_1}{k_1} \\ 
\yb\lra\fbox{$k_2$}\lra\fbox{Estimation}\lra\wh{\xb}_2  
\end{array}
\]
\centerline{A scale invariant estimation method}
}}}

\medskip\noindent 
To do this let us note  
\begin{itemize}
\item $\thetab$ all the unknown
parameters defining our measuring system (noise variance $\sigma^2$ and 
the prior law parameters for example), 

\item $p_1(\xb_1|\yb_1;\thetab_1)$ and $p_k(\xb_k|\yb_k;\thetab_k)$ the
two expressions of the posterior law for scale 1 and for scale $k$ with
\[
\xb_k=k \xb_1, \quad \yb_k=k \yb_1.
\]
\end{itemize}
Then, what we need is the following:
\begin{equation}
\exists \thetab_k= f(\thetab_1, k) \,|\, 
\forall k>0, \forall \xb_1,\yb_1,\quad  
p_k\left(\xb_k | \yb_k; \thetab_k\right) 
= \frac{1}{k^n} \, p_1(\xb_1|\yb_1;\thetab_1),
\end{equation}
which means that the functional form of the posterior law remains unchanged
when the  measurement's scale is changed. Only we have to modify the
parameters  $\thetab_k=f(\thetab_1,k)$ which is only a function of
$\thetab_1$ and the scale factor $k$.

However, not all estimators based on this posterior will be scale invariant. 
The cost function must also have some property to obtain a scale invariant
estimator. So, the main result of this paper can be resumed in the following
theorem:

\medskip\noindent{\bf Theorem:~} 
If $\exists \thetab_k=f(\thetab_1,k) \,|\, \forall k>0, \forall \xb_1,\yb_1,$
\[
p_k\left(\xb_k | \yb_k; \thetab_k\right) =
\frac{1}{k^n} \, p_1(\xb_1|\yb_1;\thetab_1), 
\]
then any bayesian estimator with a cost function $C(\wh{\xb},\xb)$ satisfying: 
\[
C(\wh{\xb}_k,\xb_k)=a_k + b_k C(\wh{\xb},\xb), 
\]
is a scale invariant estimator, \ie;
\[
\wh{\xb}_k(\yb_k;\thetab_k) = k \wh{\xb}_1(\yb_1;\thetab_1).
\]

\smallskip\noindent{\bf Proof:~}
In fact, it is easy to see the following:
\[
\begin{array}{ll}
\wh{\xb}_k(\yb_k;\thetab_k)
&=\disp{\argmins{\zb_k}{\intg C(\zb_k,\xb_k) 
   p_k(\xb_k|\yb_k;\thetab_k) \d{\xb_k}}}\\ 
&=\disp{k \, \argmins{\zb_1}{\intg [b_k C(\zb_1,\xb_1)+a_k] 
   \frac{1}{k^n} \, p_1(\xb_1|\yb_1;\thetab_1) k^n \d{\xb_1}}}\\      
&=\disp{k \, \argmins{\zb_1}{b_k \intg C(\zb_1,\xb_1)
   p_1(\xb_1|\yb_1;\thetab_1) \d{\xb_1} + a_k}} \\ 
&=\disp{k \, \argmins{\zb_1}{\intg C(\zb_1,\xb_1) 
  p_1(\xb_1|\yb_1;\thetab_1) \d{\xb_1}}} 
\\  
&= k \, \wh{\xb}_1(\yb_1;\thetab_1)
\end{array}
\]
 
Note the great significance of this result, even if the estimateur 
$\wh{\xb}(\yb;\thetab)$ is a nonlinear function of the observations $\yb$ 
it stays scale invariant.

Now, the task is to search for a large familly of probability laws $p(\xb)$
and $p(\yb|\xb)$ in a manner that the posterior law $p(\xb|\yb)$
remains scale invariant. We propose to do this search in the generalized
exponential familly for two reasons: 
\begin{itemize}
\item First the generalized exponential probability density
functions form a very rich one, and 

\item Second, they can be considered as the
maximum entropy prior laws subject to a finite number of constraints 
(linear or nonlinear).
\end{itemize}

Noting also that if $p(\xb)$ and $p(\yb|\xb)$ are scale invariant then the
posterior  $p(\xb|\yb)$ is also scale invariant and that there is a symmetry
for $p(\xb)$  and $p(\yb|\xb)$, so that it is only necessary to find the
scale invariance conditions for one of them. In the following, without loss of
generality, we consider the case where $p(\yb|\xb)$ is Gaussian:
\begin{equation} \label{eq:py}
p(\yb|\xb;\sigma^2)\propto \expf{-\chi^2(\xb,\yb;\sigma^2)},
\quad\hbox{with~}
\chi^2(\xb,\yb;\sigma^2)=\frac{1}{2\sigma^2} \, 
 [\yb-\Hb \xb]^t [\yb-\Hb \xb],
\end{equation}  
and find the conditions for $p(\xb)$ to be scale invariant. 
We choose the generalized exponential $pdf$'s for $p(\xb)$, \ie;
\begin{equation} \label{eq:px}
p(\xb;\lambdab)\propto \expf{-\sum_{i=1}^r \lambda_i \phi_i(\xb)}, 
\end{equation}  
and find the conditions on the functions $\phi_i(\xb)$ for which $p(\xb)$ is
scale invariant.

Note that these laws can be considered as the maximum entropy prior laws if
our prior knowledge is:
\begin{itemize}
\item What we know about $\xb$ is:
\[
\esp{\phi_i(\xb)}=d_i, \quad i=1,\cdots,r,
\]

\item and what we know about the noise $\bb$ is:
\[
\left\{\begin{array}{l}
\esp{\bb}=0, \\ 
\esp{\bb \bb^t}=\Rb_b=\sigma^2 \Ib,
\end{array}
\right.
\]
where $\Rb_b$ is the covariance matrix of $\bb$. 
\end{itemize}
Now, using the equations (\ref{eq:py}) and (\ref{eq:px}) and noting by 
$\thetab=(\sigma^2,\lambda_1,\cdots,\lambda_r)$, by 
$\lambdab=(\lambda_1,\cdots,\lambda_r)$, and by 
$\phib(\xb)=(\phi_1(\xb),\cdots,\phi_r(\xb))$, we have 
\begin{equation}
p(\xb|\yb;\thetab)\propto 
\expf{-\chi^2(\xb,\yb;\sigma^2)-\lambdab^t \phib(\xb)}, 
\end{equation}  
and the scale invariance condition becomes:
\[
\forall k>0, \forall \xb_1,\yb_1,\quad 
\chi^2_k(\xb_k,\yb_k;\sigma_k^2)+\lambdab_k^t \phib(\xb_k)=
\chi^2_1(\xb_1,\yb_1;\sigma_1^2)+\lambdab_1^t \phib(\xb_1)+cte.   
\]
But with the Gaussian choice for the noise $pdf$ we have
\[
\forall k>0, \forall \xb_1,\yb_1,~ 
\chi^2_k(\xb_k,\yb_k;\sigma_k^2)
=\frac{1}{2\sigma_k^2}\, ||\yb_k-\Hb \xb_k||^2 \\ 
=\frac{1}{2 k^2 \sigma_1^2} \, k^2 \, ||\yb_1-\Hb \xb_1||^2 \\
=\chi^2_1(\xb_1,\yb_1;\sigma_1^2), 
\]
and so the condition becomes 
\begin{equation} \label{eq:phi}
\forall k>0, \forall \xb,\quad   
\lambdab_k^t \phib(\xb_k)=\lambdab_1^t \phib(\xb_1) + cte,
\end{equation}
or equivalently, 
\[
p_k(\xb_k;\lambdab_k) =\frac{1}{k^n} \, p_1(\xb_1;\lambdab_1)
\quad\hbox{~with~}\quad 
\lambdab_k=f(\lambdab_1, k). 
\]
Thus, in the case of centered Gaussian $pdf$ for the noise, to have a scale
invariant posterior law it is sufficient to have a scale invariant prior law. 

Now, assuming interchangeable (independent) pixels, \ie;
\begin{equation}
p(\xb;\lambdab)=\expf{\lambda_0+\sum_{i=1}^r \lambdab_i \phi_i(\xb)} 
= \prod_{j=1}^N p(x_j;\lambdab), 
\end{equation}
or equivalently, 
\begin{equation}
\phi_i(\xb)=\sum_{j=1}^N \phi_i(x_j)
\end{equation}
we have to find the conditions on the scalar functions $\phi_i(x)$ of 
scalar variables $x$ who satisfy the equation (\ref{eq:phi}) or 
equivalently
\begin{equation}
\forall k>0, \forall x,\quad 
\sum_{i=1}^r \lambda_{i}(k) \, \phi_i(k x) 
= \sum_{i=1}^r \lambda_{i}(1) \, \phi_i(x)+cte  
\end{equation}
We have shown (see appendix) that, 
the functions ${\phi_i(x)}$ which satisfy these conditions are all either 
the powers of $x$ or the powers of $\ln x$ or a multiplication of them. 
The general expressions for these functions are:
\begin{equation}
\phi(x)=\sum_{m=1}^M 
\left( \sum_{n=0}^{N_m-1} c_{mn} (\ln x)^n \right) x^{\alpha_m}
+ \sum_{n=0}^{N_0} c_{0n} (\ln x)^n, 
\quad\hbox{with~}
M\le r 
\hbox{~and~}
\sum_{m=0}^M N_m = r
\end{equation}
where $M$ and $N_m$ are integer numbers, and $c_{mn}$, $c_{0n}$ and
$\alpha_m$ are  real numbers. For a geometrical interpretation and more
details see appendix.  The following examples show some special and
interesting cases.

\medskip\noindent {\bf One parameter laws:}
Consider the case of $r=1$. In this case we have
\begin{equation}
p(x;\lambda)\propto \expf{-\lambda \phi(x)}.
\end{equation}
Applying the general rule with 
\[
r=1\lra 
\left\{\begin{array}{ll} M=0, N_0=1,          &\lra c_{00} + c_{01} \ln x \\
                  M=1, N_0=0, N_1=1,  &\lra c_{00} + c_{10} x^{\alpha_1}
\end{array}\right.
\]
we find that the only functions who satisfy these conditions are:  
\begin{equation}\label{oneparam}
\biggl\{\phi(x)\biggr\}=
\biggl\{x^{\alpha}, \ln x\biggr\} 
\end{equation}
where $\alpha$ is a real number. 
There isc two interesting special cases: 

\begin{itemize}
\item $\phi(x)=x^{\alpha}$, resulting to:\quad
$p(x)\propto \expf{-\lambda x^{\alpha}},\, \alpha>0, \lambda>0,$ 
which is a generalized Gaussian $pdf$, and 

\item $\phi(x)=\ln x$, resulting to:\quad 
$p(x)\propto \expf{-\lambda \ln x},$   
which is a special case of the Beta $pdf$.
\end{itemize}
Note that the famous {\em entropic} prior law: \quad 
$p(x)\propto \expf{-\lambda x \ln x}$  
of Gull and Skilling \cite{Skilling-88,Gull-84} does not verify the 
scale invariance property. But, if we add one more parameter 
\[
p(x)\propto \expf{-\lambda x \ln x + \mu x},
\] 
then, it will satisfy this condition as we can see in the next section.  

\medskip\noindent{\bf Two parameters laws:}
This is the case where $r=2$ and we have:
\begin{equation}
p(x;\lambda)\propto \expf{-\lambda \phi_1(x) - \mu \phi_2(x)},
\end{equation}
and applying the general rule:
\[
r=2\lra 
\left\{\begin{array}{ll} 
    M=2,N_0=0,N_1=1,N_2=1, 
    &\lra c_{00}+c_{10} x^{\alpha_1}+c_{20} x^{\alpha_2}  \\
    M=1,N_0=0,N_1=2, 
    &\lra c_{00}+c_{10} x^{\alpha_1}+c_{11} x^{\alpha_1} \ln x  \\
    M=1,N_0=1,N_1=1, 
    &\lra c_{00}+c_{10} x^{\alpha_1}+c_{01} \ln x  \\
    M=0,N_0=2, 
    &\lra c_{00}+c_{01} \ln x+c_{02} \ln^2 x 
\end{array}\right.
\]
we see that in this case the only functions $(\phi_1, \phi_2)$
which satisfy these conditions are:
 
\begin{equation}\label{twoparam1}
\begin{array}{ll}
\biggl\{\left(\phi_1(x),\phi_2(x)\right)\biggr\}= &
\biggl\{
(x^{\alpha_1},x^{\alpha_2}), (x^{\alpha_1},x^{\alpha_1} \ln x), 
(x^{\alpha_1},\ln x), (\ln x ,\ln ^2 x)
\biggr\}
\end{array}
\end{equation}
where $\alpha_1$ and $\alpha_2$ are two real numbers. 
Special cases are obtained when we choose \\ 
$\phi_2(x)=x$, the only possible functions for $\phi_1(x)$ are then:  
\begin{equation} \label{twoparam2}
\left\{x^{\alpha}, \ln x, x \ln x\right\}.
\end{equation}
and we have the following interesting cases:

\begin{itemize}
\item $\phi_1(x)=x^2$, resulting to: \quad
$
p(x)\propto \expf{-\lambda x^2-\mu x} 
     \propto \expf{-\lambda \left(x+\frac{\mu}{2\lambda}\right)^2},
$
which is a Gaussian $pdf$ 
${\cal N}\left(m=\frac{-\mu}{\lambda},\sigma^2=\frac{1}{2\lambda}\right)$.
 
\item $\phi_1(x)=\ln x$, resulting to:\quad
$
p(x)\propto \expf{-\lambda \ln x - \mu x} 
= x^{-\lambda} \expf{-\mu x},
$
which is the Gamma $pdf$, and finally,

\item $\phi_1(x)=x\ln x$, resulting to:\quad
$
p(x)\propto \expf{- \lambda x\ln x - \mu x}. 
$
which is known as the {\it entropic\/} $pdf$.
\end{itemize}

\noindent{\bf Three parameters laws:}
This is the case where $r=3$. Once more applying the general rule we find:
\[
r=3\ra 
\left\{\begin{array}{ll} 
M=3,N_0=0,N_1=1,N_2=1,N_3=1, 
 &\ra c_{00}+c_{10} x^{\alpha_1}+c_{20} x^{\alpha_2}+c_{30} x^{\alpha_3}
\\ 
M=2,N_0=0,N_1=1,N_2=2, 
 &\ra c_{00}+c_{10} x^{\alpha_1}+c_{20} x^{\alpha_2}
+c_{21} x^{\alpha_2} \ln x  
\\ 
M=2,N_0=1,N_1=1,N_2=1, 
 &\ra c_{00}+c_{01} \ln x+c_{10} x^{\alpha_1}+c_{20} x^{\alpha_2}  
\\
M=1,N_0=0,N_1=3, 
 &\ra c_{00}+c_{10} x^{\alpha_1}+c_{11} x^{\alpha_1} \ln x
+c_{12} x^{\alpha_1} \ln^2 x 
\\
M=1,N_0=1,N_1=2, 
 &\ra c_{00}+c_{01} \ln x+c_{10} x^{\alpha_1}+c_{11} x^{\alpha_1} \ln x  
\\
M=1,N_0=2,N_1=1, 
 &\ra c_{00}+c_{01} \ln x+c_{02} \ln^2 x+c_{10} x^{\alpha_1} \\
M=0,N_0=3, 
 &\ra c_{00}+c_{01} \ln x+c_{02} \ln^2 x+c_{03} \ln^3 x 
\end{array}\right.
\] 
which means:
\begin{equation}
\begin{array}{llll}
\biggl\{\left(\phi_1(x),\phi_2(x),\phi_3(x)\right)\biggr\}
 =\biggl\{ & 
 (x^{\alpha_1}, x^{\alpha_2}, x^{\alpha_3}), \,  
 (x^{\alpha_1}, x^{\alpha_2},       \ln x), \, 
 (x^{\alpha_1}, x^{\alpha_1}\ln x, x^{\alpha_1} \ln^2 x), \\      
&(x^{\alpha_1}, x^{\alpha_1}\ln x, \ln x), \,         
 (x^{\alpha_1}, x^{\alpha_2},      x^{\alpha_2} \ln x), \,                      
 (x^{\alpha_1}, \ln x,               \ln ^2 x), \\  
&\hspace{6.5cm}(\ln x ,        \ln ^2 x           , \ln ^3 x) \biggr\}
\end{array}
\end{equation}
where $\alpha_1$, $\alpha_2$ and $\alpha_3$ are three real numbers. 

\section{Proposed method}
The general procedure of the inversion method we propose can be resumed 
as follows: 
\begin{itemize}
\item Choose a set of functions $\phi_i(x)$ between the possibles ones 
described in the last section and assign the prior $p(\xb)$. In many 
imaging applications we proposed and used successfully the following two 
parameters one:
\[
p(\xb;\lambdab)\propto
\expf{-\lambda_1 H(\xb)-\lambda_2 S(\xb)},
\quad\hbox{with~}
H(\xb)=\sum_{j=1}^N \phi_1(x_j), 
\hbox{~and~} 
S(\xb)=\sum_{j=1}^N \phi_2(x_j)
\]
where $\phi_1(x)$ and $\phi_2(x)$ choosed between the possible ones in 
(\ref{twoparam1}) or (\ref{twoparam2}). 

\item When what we know about the noise $\bb$ is only its covariance matrix 
$\esp{\bb\bb^t}=\Rb_b=\sigma_b^2\Ib$, then using the maximum entropy
principle we have: 
\[
\pyx\propto \expf{-\frac{1}{2} Q(\xb)},
\quad\hbox{with~}
Q(\xb)=(\yb-\Ab\xb)^t {\Rb_b}^{-1} (\yb-\Ab\xb).
\]
We may note that $\pyx$ is also a scale invariant probability law.

\item Using the Bayes' rule and MAP estimator the solution is determined by 
\[
\wh{\xb}=\argmaxs{\xb}{\pxy}=\argmins{\xb}{J(\xb)},
\quad\hbox{with~}
J(\xb)=Q(\xb)+\lambda_1 H(\xb)+\lambda_2 S(\xb).
\]
Note here also that, for the cases where one of the functions $\phi_1(x)$ or 
$\phi_2(x)$ is a logarithmic function of $x$, we have to constraint its range 
to the positive real axis, and we have to solve the following optimization
problem  
\[
\wh{\xb}=\argmaxs{\xb>0}{\pxy}=\argmins{\xb>0}{J(\xb)}.
\]
This optimization is achieved by a modified conjugate gradients method. 

\item The choice of the functions $\phi_i(x)$ and the determination of the
parameters  $(\lambda_1,\lambda_2)$ in the first step is still an open
problem. 

In imaging applications we propose to do this choice from our prior knowledge
on the nature of interested quantity (physics of the application). For
example, if the object $\xb$ is a real quantity equally distributed on the
positive and the negative reals then a Gaussian prior, \ie;  $(\phi_1(x)=x,
\phi_2(x)=x^2)$ is convenient. But, if the object $\xb$ is a positive
quantity or if we know that it represents small extent, bright and sharp
objects on a nearly black background (images in radio astronomy, for
example), then we may choose  $(\phi_1(x)=x, \phi_2(x)=\ln x)$, or
$(\phi_1(x)=x, \phi_2(x)=x \ln x)$ which are the priors with longer tails
than the Gaussian or truncated Gaussian one. 

When the choice of the functions $(\phi_1(x), \phi_2(x))$ is done, we still
have to  determine the hyperparameters $(\lambda_1,\lambda_2)$. For this two
main approaches have been proposed. The first is based on the generalized
maximum likelihood (GML) which tries to estimate simultaneously the
parameters $\xb$ and  the hyperparameters $\thetab=(\lambda_1,\lambda_2)$
by 
\begin{equation} 
(\wh{\xb},\wh{\thetab})= \argmaxs{(\xb,\thetab)}{p(\xb,\yb;\thetab)} 
                 = \argmaxs{(\xb,\thetab)}{p(\yb|\xb) \, p(\xb;\thetab)},
\end{equation}
and the second is based on the marginalization (MML), in which 
the hyperparameters $\thetab$ are estimated first by 
\begin{equation}
\wh{\thetab}= \argmaxs{\thetab}{p(\yb;\thetab)
         =\intg p(\xb,\yb;\thetab) \d{\xb}} \\
         = \argmaxs{\thetab}{\intg p(\yb|\xb)\, p(\xb;\thetab) \d{\xb}},
\end{equation}
and then used for the estimation of $\xb$:
\begin{equation}
\wh{\xb}= \argmaxs{\xb}{p(\xb|\yb;\wh{\thetab})} 
     = \argmaxs{\xb}{p(\yb|\xb)\, p(\xb|\wh{\thetab})}.
\end{equation}

What is important here is that both methods preserve the scale invariant
property. For practical applications we have recently proposed and used a
method based on the generalized maximum likelihood \cite{AMD93a,AMD93b}
which has been successfully used in many signal and image reconstruction and
restoration problems as we mentionned in the introduction \cite{Nguyen-92}.
\end{itemize}

\section{Conclusions}
Excepted the Gaussian case where all the Bayesian estimators are linear
functions of the observed data, in general, the Bayesian estimators are 
{\em nonlinear} functions of the data. 
When dealing with linear inverse problems linearity is sometimes a too strong
property, while {\em scale invariance} often remains a desirable property.   
In this paper we discussed and proposed a familly of generalized exponential
probability distributions for the direct probabilities (the prior $p(\xb)$ and
the likelihood $p(\yb|\xb)$), for which the posterior $p(\xb|\yb)$, and,
consequently, the main posterior estimators are scale invariant. 
Among many properties, generalized exponential can be considered as the
maximum entropy probability distributions subject to the knowledge of a finite
set of expectation values of some knwon functions. 

\renewcommand{\theequation}{A.\arabic{equation}}
\setcounter{section}{-1}
\setcounter{equation}{0}
\appendix
\section{Appendix: General case}
We want to find the solutions of the following equation:

\begin{equation} \label{app1}
\forall k>0, \forall x,\quad
\disp{\sum_{i=1}^r \lambda_{i}(k) \phi_i(kx) 
= \sum_{i=1}^r \lambda_{i}(1) \phi_i(x) +\beta(k)}
\end{equation}
Making the following changes of variables and notations
\[
1/k=\tilde k, \, k x=\tilde x, \, 
\lambda_{i}(k)=\tilde{\lambda}_{i}(\tilde k), \hbox{~~and~~} 
\beta_{i}(k)=\tilde{\beta}_{i}(\tilde k),
\]
equation (\ref{app1}) becomes
\[
\sum_{i=1}^r \tilde{\lambda}_{i}(\tilde k) \phi_i(\tilde x) 
= \sum_{i=1}^r \tilde{\lambda}_{i}(1) \phi_i(\tilde k \tilde x)
+\tilde{\beta}(\tilde k)
\]
For convenience sake, we will drop out the tilde~$\tilde{~}$, and note 
$\lambda_{i}(1)=\lambda_{i}$, so that we can write  
\[
\sum_{i=1}^r \lambda_{i}(k) \phi_i(x) 
= \sum_{i=1}^r \lambda_{i} \phi_i(kx)+\beta(k)
\]
Noting  
\[
S(x)=\sum_{i=1}^r \lambda_{i} \phi_i(x),
\quad\hbox{and so~~~}
S(kx)=\sum_{i=1}^r \lambda_{i} \phi_i(kx)
\]
we have

\begin{equation} \label{app2}
\sum_{i=1}^r \lambda_{i}(k) \phi_i(x)=S(kx)+\beta(k)  
\end{equation}
Deriving $r-1$ times this equation with respect to $k$ we obtain
\begin{equation}\label{app3}
\begin{array}{ll}
 \disp{\sum_{i=1}^r \lambda'_{i}(k) \phi_i(x)}
&\disp{= x \, S'(kx)+\beta'(k)}                    \\
 \disp{\sum_{i=1}^r \lambda''_{i}(k) \phi_i(x)}
&\disp{= x^2 \, S''(kx)+\beta''(k)}                \\
 \vdots  & \vdots                                  \\
\disp{\sum_{i=1}^r \lambda^{(r-1)}_{i}(k) \phi_i(x)}
&\disp{= x^{r-1} S^{(r-1)}(kx)+\beta^{(r-1)}(k)}
\end{array}
\end{equation}
Combinig equations (\ref{app2}) and (\ref{app3}) in matrix form we have
\begin{equation}\label{app4}
\pmatrix{
\lambda_1(k)  & \cdots &\lambda_r(k)\cr 
\lambda'_1(k) & \cdots &\lambda'_r(k)\cr 
\lambda''_1(k) & \cdots &\lambda''_r(k)\cr 
\vdots        & \cdots &\vdots      \cr
\lambda^{(r-1)}_1(k) & \cdots &\lambda^{(r-1)}_r(k)}
\, 
\pmatrix{
\phi_1(x) \cr
\phi_2(x) \cr
\phi_3(x) \cr
\vdots     \cr
\phi_{r}(x)}
=\pmatrix{
S(kx)+\beta(k)         \cr 
x S'(kx)+\beta'(k)     \cr 
x^2 S''(kx)+\beta''(k) \cr 
\vdots                   \cr 
x^{r-1} S^{(r-1)}(kx)+\beta^{(r-1)}(k)}
\end{equation}

If this matrix equation can be inverted, this means that any 
function $\phi_i(x)$ is a linear combination of $S(kx)+\beta(k)$ 
and its $(r-1)$ derivatives with respect to $k$:
\begin{equation}\label{app5}
\phi_i(x)=\sum_{i=0}^r \eta_{i}(k) \left[ x^{(i-1)} S^{(i-1)}(kx) 
+ \beta^{(i-1)}(k) \right],
\end{equation}
and if this is not the case, this means that there exists an 
interval for $k$, for which some of the functions $\lambda_i(k)$ 
are linear combinations of the others \cite{Bass-68}. 
In this case let us show that we will go back to the situation of the 
problem of lower order $r$. Let us to assume that the last column of the 
matrix is a linear combination of the others, \ie; 
\[
\lambda_r(k)=\sum_{i=1}^{r-1} \gamma_i \lambda_i(k).
\]
Putting this in the equation (\ref{app1}) will give
\[
\sum_{i=1}^{r-1} \lambda_{i}(k) \phi_i(kx) 
 + \left[\sum_{i=1}^{r-1} \gamma_i \lambda_{i}(k)\right] \phi_r(kx) 
=
\sum_{i=1}^{r-1} \lambda_{i}(1) \phi_i(x) +\beta(k)  
 +\left[\sum_{i=1}^{r-1} \gamma_i \lambda_{i}(1) \right] \phi_r(x)
\]
and noting 
$\psi_i(x)= \phi_i(x)+ \gamma_i \phi_r(x)$ and 
$\psi_i(kx)= \phi_i(kx)+ \gamma_i \phi_r(kx)$ we obtain
\[
\sum_{i=1}^{r-1} \lambda_{i}(k) \psi_i(kx) = 
\sum_{i=1}^r \lambda_{i}(1) \psi_i(x) +\beta(k)
\]
which is an equation in the same form of (\ref{app1}), 
but of lower order. 

Deriving now both parts of the equation (\ref{app5}) 
with respect to $k$ and noting $kx=u$ we obtain
\begin{equation}\label{app6}
\sum_{i=0}^r a_{i} \, u^i S^{i}(u)=a  
\end{equation}
This is the general expression of a $r$th order Euler--Cauchy 
differential equation \cite{Angot-82,Bass-68} which is classically 
solved through the change of variable $u=e^x$, and one can find the 
general expression of its solution in the following form:
\begin{equation} \label{app7}
S(x)=\sum_{m=1}^M \left( \sum_{n=0}^{N_m-1} 
c_{mn} (\ln x)^n \right) x^{\alpha_m}
+ \sum_{n=0}^{N_0} c_{0n} (\ln x)^n
\quad\hbox{with~} M=0,\cdots r, 
\hbox{~~and~~}
\sum_{m=0}^M N_m = r
\end{equation}
where $M$ and $N_m$ are integer numbers, and $c_{mn}$, 
$c_{0n}$ and $\alpha_m$ are real numbers. 
In fact the most general solution also incorporate terms 
of the form
\[
\left[\sum_n (\ln x)^n \left(\alpha_n \cos(\ln x) 
+ \beta_n \sin(\ln x) \right)\right] 
\, x^{d} 
\]
derived from complex $\alpha_m$ and $c_{mn}$. 
But we will not consider these terms because 
the resulting $pdf$'s have oscillatory behavior around zero. 

One can give a geometric interpretation of the solutions given in 
(\ref{app7}).   
For any given order $r$ make a $(r+1) \times (r+1)$ table in the form
\[
\begin{array}{c|c|c|c|c|c|} \cline{1-2}
\ln^r x   & ~~      \\ \cline{1-3}
{\vdots}     & ~~    & ~~   \\ \cline{1-4} 
\ln^2 x   & ~~    & ~~ & ~~   \\ \cline{1-5}
\ln x      & ~~    & ~~ & ~~ & ~~  \\ \cline{1-6}  
1           & \times & ~~ & ~~ & ~~ & ~~ \\ \cline{1-6} 
           & 1       &x^{\alpha_1}&x^{\alpha_2} &\cdots &x^{\alpha_r}
\end{array}
\]
and let $r$ mass points fall down into the columns. To each filled box 
is assigned a function $\phi_i(x)$ by multiplying the corresponding 
powers of $x$ and $\ln x$ on the same line and the same column. 
To illustrate this, we give in the following the three first cases: 
\[
\begin{array}{|c|c|c|}  \cline{1-3}\cline{1-3}
\hspace{12mm} \hbox{Case $r=1$:} \hspace{12mm} & 
\hspace{12mm} \hbox{Case $r=2$:} \hspace{12mm} & 
\hbox{Case $r=3$:} \\ \cline{1-3} &&\\ 
\begin{array}[b]{c|c|c|} \cline{1-2}
\ln x      & b            \\ \cline{1-3}  
1           & \times  & a       \\ \cline{1-3}  
            &1        & x^{\alpha_1} 
\end{array}
&
\begin{array}[b]{c|r|r|r|} \cline{1-2}
\ln^2 x   &  d      \\ \cline{1-3}
\ln x      & bd     &   c  \\ \cline{1-4}  
1           & \times & abc & a    \\ \cline{1-4}  
            & 1      & x^{\alpha_1} & x^{\alpha_2} 
\end{array}
&
\begin{array}[b]{c|r|r|r|r|} \cline{1-2}
\ln^3 x   &    g     \\ \cline{1-3}
\ln^2 x   &   fg    &   c~    \\ \cline{1-4} 
\ln x      & bdfg    &    dc~  & ~e   \\ \cline{1-5}  
1           & \times & abcdef & abe & a    \\ \cline{1-5}  
            & 1       & x^{\alpha_1} & x^{\alpha_2} & x^{\alpha_3} 
\end{array}
\\ &&\\ \cline{1-3} &&\\
\begin{array}[t]{l|c}
 & \phi(x)      \\ \hline
a& x^{\alpha_1} \\ 
b& \ln x 
\end{array} 
&
\begin{array}[t]{l|cc}
 & \phi_1(x)  & \phi_2(x)            \\ \hline
a& x^{\alpha_1} & x^{\alpha_2}      \\ 
b& x^{\alpha_1} & \ln x               \\ 
c& x^{\alpha_1} & x^{\alpha_1} \ln x \\ 
d& \ln x         & \ln ^2 x
\end{array}
&
\begin{array}[t]{l|ccc}
 & \phi_1(x)    &   \phi_2(x)   &   \phi_3(x)      \\ \hline
a& x^{\alpha_1} & x^{\alpha_2} &   x^{\alpha_3}   \\ 
b& x^{\alpha_1} & x^{\alpha_2} &   \ln x            \\ 
c& x^{\alpha_1} & x^{\alpha_1}\ln x & x^{\alpha_1} \ln^2 x \\ 
d& x^{\alpha_1} & x^{\alpha_1}\ln x & \ln x                 \\ 
e& x^{\alpha_1} & x^{\alpha_2}       & x^{\alpha_2} \ln x  \\ 
f& x^{\alpha_1} & \ln x               &               \ln ^2 x \\ 
g& \ln x         &         \ln ^2 x            & \ln ^3 x   
\end{array}  
\\ &&\\ \cline{1-3}\cline{1-3}
\end{array}
\]

\newpage

\end{document}